# Balancing Privacy, Utility, and Safety in LLM Alignment through Preference Optimization

Dishu Yang*†
*Khoury College of Computer Sciences*
*Northeastern University*
San Jose, USA
yang.dis@northeastern.edu
*Corresponding author

Jingjing (May) Liu†
*Department of Electrical Engineering and Computer Sciences*
*University of California, Berkeley*
Berkeley, USA
ORCID: 0009-0009-6936-4017

Jize Li
*Department of Economics*
*Boston University*
Boston, USA
jizel@bu.edu

†Equal contribution as a first author

*Abstract*— **Preference optimization is widely used to align large language models with human preferences, but preference-data composition may also influence privacy-relevant memorization. We examine whether adding synthetic privacy-preference pairs to Direct Preference Optimization (DPO) is associated with lower canary-based memorization signals without modifying the objective or introducing a formal privacy mechanism. We propose Privacy-Pressure Preference Mixing (P3M), a data-composition protocol that varies the amount of privacy-preference data while keeping helpfulness and harmlessness preference data fixed. We evaluate a non-privacy Baseline and privacy-mixing ratios of 0.5, 1.0, and 2.0 using Gemma 3 270M-IT across five random seeds and validate the same four conditions using 4-bit-quantized Gemma 2 2B-IT across three seeds. Overall, under the tested conditions, privacy-preference mixing is associated with lower mean canary suffix log-likelihood proxy values across both model settings and lower aggregate membership-inference attack performance relative to the Baseline in the mixed-source 2B evaluation. Specifically, across the privacy-aware 2B configurations, the mean area under the receiver operating characteristic curve (AUROC) ranges from 0.596 to 0.629, and the mean area under the precision-recall curve (AUPRC) ranges from 0.541 to 0.575, compared with 0.804 and 0.790, respectively, for the Baseline. However, the reduction in membership distinguishability does not hold uniformly across data sources. Moreover, the relationship between the privacy ratio and harmlessness preference accuracy varies by model setting, whereas helpfulness preference accuracy remains broadly stable. These findings suggest that P3M should be viewed as a lightweight empirical protocol for examining privacy-utility-safety trade-offs rather than as a formal privacy guarantee or a defense against extraction attacks.**

**Keywords— DPO, privacy-aware LLM alignment, preference mixing, LoRA, trustworthy AI, privacy-utility-safety trade-off**

## I. Introduction

Preference-based alignment methods train language models using human judgments over alternative outputs. In Reinforcement Learning from Human Feedback (RLHF), models are fine-tuned from human demonstrations or rankings to better follow instructions and align with human intent [1,2,3,4]. Direct Preference Optimization (DPO) bypasses the reward-modeling step by directly optimizing the language model on preference data [5]. However, alignment data may contain prompts, responses, preference labels, and user-derived context. Such data can be vulnerable to membership-inference attacks [6], and memorization may surface or persist across RLHF and direct preference-learning pipelines, raising privacy concerns when sensitive examples are later regurgitated [7].

Recent privacy-preserving alignment work addresses related risks through differential privacy for RLHF-style alignment [8], privacy-preserving synthetic instructions [9], federated RLHF [10], and differentially private synthetic preference data [12]. These approaches rely on formal privacy mechanisms, decentralized training, or additional synthesis pipelines. This paper studies a lighter and more practically accessible question: when the DPO objective, LoRA setup, base model, and non-privacy data are fixed within each experiment, how do helpfulness, harmlessness, and memorization-related signals change as the amount of privacy-oriented preference data varies? We further evaluate the same data-composition conditions in a larger-model setting and conduct a complementary membership-inference analysis.

We propose Privacy-Pressure Preference Mixing (P3M), a data-composition protocol that treats privacy pressure as an experimental variable rather than as a new objective. P3M adds synthetic privacy-preference pairs to a standard helpfulness-and-harmlessness preference mixture. In each privacy pair, the chosen response does not disclose a synthetic private identifier, whereas the rejected response reveals it. Across configurations, we vary privacy pressure by changing the amount of privacy-preference data in the DPO training pool while keeping the helpfulness and harmlessness data fixed.

Our controlled, resource-bounded evaluation compares a non-privacy Baseline with three privacy-mixing ratios (0.5, 1.0, and 2.0) using Gemma 3 270M-IT across five random seeds and 4-bit Gemma 2 2B-IT across three seeds. Both experiments use LoRA-based DPO and evaluate helpfulness preference accuracy, harmlessness preference accuracy, and a canary suffix log-likelihood proxy following the broader canary-based memorization methodology [13]. For the 2B validation, we additionally adapt PREMIA, a reference-based membership-inference framework for preference data [6]. We report attack performance using the area under the receiver operating characteristic curve (AUROC) and the area under the precision-recall curve (AUPRC). Together, the canary-based proxy and PREMIA-based evaluation provide complementary empirical signals of memorization and membership distinguishability, but they do not provide a formal privacy guarantee or demonstrate resistance to extraction attacks.

Across both model settings, all privacy-aware configurations yield lower mean canary suffix log-likelihood proxy values than the corresponding Baselines. In the 2B validation, they also yield lower aggregate AUROC and AUPRC on the constructed mixed-source attack sets, although source-stratified results do not show uniform reductions across helpfulness, harmlessness, and privacy-preference tuples. Helpfulness preference accuracy remains broadly stable, whereas harmlessness preference accuracy decreases monotonically with the privacy ratio in the 270M experiment but not in the 2B validation. These results highlight the importance of jointly evaluating privacy, utility, and safety outcomes.

This paper makes three contributions. First, it introduces Privacy-Pressure Preference Mixing (P3M), a lightweight protocol for varying privacy-oriented preference data within DPO. Second, it evaluates the same four data-composition conditions using Gemma 3 270M-IT across five random seeds and 4-bit Gemma 2 2B-IT across three random seeds, jointly measuring helpfulness preference accuracy, harmlessness preference accuracy, and the canary suffix log-likelihood proxy. Third, it complements the canary-based evaluation with a PREMIA-based membership-inference evaluation on the 2B model and reports attack performance at both the aggregate mixed-source and source-stratified levels, showing that the reduction in aggregate attack performance observed for the privacy-aware configurations relative to the Baseline does not hold uniformly across helpfulness, harmlessness, and privacy-preference tuples.

## II. Related Work

### A. Preference Optimization for LLM Alignment

Preference-based learning uses human comparisons to learn reward functions [1] and has subsequently been applied to summarization [2], instruction following [3], and helpful-and-harmless assistant training [4]. The helpful-and-harmless setting motivates the utility-safety framing adopted in this study, but privacy exposure is not a primary training or evaluation target in these alignment applications. Conventional RLHF typically involves separate reward-model training and policy optimization, whereas DPO directly optimizes the policy from preference pairs without fitting a separate reward model [5]. This simpler training structure makes DPO suitable for controlled studies of preference-data composition. Building on DPO, P3M keeps the optimization objective unchanged and examines whether adding privacy-preference pairs affects memorization-related signals as well as helpfulness and harmlessness outcomes.

### B. Memorization and Privacy Leakage in Language Models

Language models may memorize rare, unique, or duplicated training sequences, creating a risk of data exposure. Canary-based testing inserts synthetic sequences and quantifies their exposure [13], while extraction attacks show that models can reproduce verbatim training examples, including personally identifiable information [15]. Memorization increases with model capacity, example duplication, and the amount of prompt context [16]. These concerns extend to alignment: preference data can be vulnerable to membership-inference attacks in DPO- and PPO-aligned models [6], while memorization can surface and persist through RLHF and direct preference-learning pipelines [7]. Accordingly, we use a canary suffix log-likelihood proxy in both model settings and a PREMIA-based membership-inference evaluation on the 2B model as complementary empirical diagnostics.

### C. Privacy-Aware Alignment and Safety Trade-offs

Privacy, utility, and safety do not necessarily improve together under preference alignment. Safe RLHF separates reward and safety objectives using distinct reward and cost models [17], while broader evaluations find uneven effects of preference alignment across trustworthiness dimensions, including privacy [18]. Privacy-aware alignment has explored the application of differential privacy to reinforcement-learning-based alignment [8], privacy-preserving synthetic instructions [9], federated RLHF [10], user-level private RLHF [11], and differentially private preference-data synthesis [12]. P3M complements these approaches by retaining DPO and treating privacy-preference mixing as a data-composition intervention evaluated across three dimensions: memorization-related signals, helpfulness, and harmlessness.

### D. Lightweight Alignment under Resource Constraints

LoRA reduces the number of trainable parameters by freezing the base model and optimizing low-rank adapters [14], while QLoRA further reduces memory usage by training LoRA adapters over a frozen 4-bit-quantized base model [19]. These methods make repeated controlled comparisons more feasible under limited compute. LoRA is used in both experiments, with 4-bit quantization applied only in the 2B validation. Both serve as computational tools rather than privacy mechanisms. The privacy intervention is the P3M data-composition protocol.

## III. Methodology

### A. Overview of the Framework

The proposed P3M framework treats privacy pressure as a configurable data-composition variable in DPO. Within each model experiment, the base model, DPO objective, LoRA configuration, and helpfulness and harmlessness pools remain fixed, while the privacy-preference data varies across mixing ratios. The resulting adapters are evaluated using helpfulness and harmlessness preference accuracy and a canary suffix log-likelihood proxy. Section III-F extends the evaluation with a larger-model validation and PREMIA-based membership inference. P3M changes data composition rather than the optimization objective and does not provide a formal privacy mechanism.

### B. Preference Data Construction

The fixed non-privacy data comprise helpfulness and harmlessness preference pairs from Anthropic HH-RLHF, representing utility and safety, respectively. Each example follows the standard preference-pair format used in DPO training:

$$\{prompt, chosen\ response, rejected\ response\} \quad (1)$$

### C. Synthetic Privacy-Preference Data and Canary Design

To examine whether adding privacy-preference data is associated with lower values of the canary suffix log-likelihood proxy, we construct synthetic preference pairs around artificial identifiers: the chosen response does not disclose the identifier, whereas the rejected response discloses it. The base privacy pool comprises 100 unique identifiers assigned to low-, medium-, and high-frequency tiers with 1 ($n = 33$), 5 ($n = 33$), and 10 ($n = 34$) nominal occurrences, respectively, yielding 538 rows. The sampling pool at privacy ratio 0.5 contains 269 rows drawn from the base pool and covers 83 identifiers, whereas the pools at privacy ratios 1.0 and 2.0 contain all 100 identifiers. None of these identifiers appears in the Baseline training pool. The evaluation covers all 100 base-pool identifiers rather than a separate held-out set. Because each adapter is trained for only 100 update steps using examples sampled from its pool, inclusion in the nominal pool does not guarantee that a given identifier appears in the realized update stream. This design provides a controlled memorization-related signal without using real personal information.

### D. Privacy Mixing Strategy and Training Setup

Privacy pressure is introduced through dataset composition. Let $D_h$ denote the helpfulness preference set, $D_s$ denote the harmlessness preference set, and $D_p$ denote the synthetic privacy-preference set. The main privacy-aware training set is constructed as:

$$D(r) = D_h \uplus D_s \uplus M(D_p, r) \quad (2)$$

where $M(D_p, r)$ samples half of, includes all of, or duplicates the privacy-preference rows for privacy mixing ratio $r \in \{0.5, 1.0, 2.0\}$, respectively. Thus, privacy pressure is implemented through data frequency rather than per-example weighting. Across conditions, we hold the base model, optimizer, DPO objective, and LoRA configuration fixed and train each adapter for 100 optimization steps. Only the amount of privacy-preference data is varied, and the nominal sampling-pool size changes accordingly.

### E. Evaluation Protocol

We evaluate the trained models across three dimensions: utility, safety, and privacy.

**Utility and safety** are measured as preference accuracy on held-out helpfulness and harmlessness sets, respectively. For each preference pair, we compute the length-normalized conditional log probability of the chosen and rejected completions given the same prompt. A prediction is correct when the chosen completion receives the higher score.

**Privacy** is empirically evaluated using the canary suffix log-likelihood proxy, defined for each canary as the average conditional log-probability per token of its hidden suffix given a prompt containing its visible prefix. Scores are first averaged within the low-, medium-, and high-frequency tiers and then averaged equally across the three tiers. Lower values indicate a weaker measured memorization-related signal. This model-dependent proxy is an empirical diagnostic, not a formal privacy guarantee or extraction test. Results are reported as the mean ± sample standard deviation across five seeds for the primary experiment and three seeds for the 2B validation.

### F. Larger-Model and Membership-Inference Validation

For validation, we repeat the four-condition comparison using 4-bit-quantized Gemma 2 2B-IT across three seeds. Table 2 summarizes the LoRA-DPO settings. Reference log-probabilities are precomputed using the frozen pre-DPO model. Because the two model settings differ in model generation and precision, the 2B experiment is a larger-model validation rather than a controlled scaling study.

Following PREMIA's target/reference probability-ratio construction [6], we use a length-normalized log-domain adaptation. For a preference tuple $z = (x, y_w, y_l)$, where $x$ denotes the prompt and $y_w$ and $y_l$ denote the chosen and rejected responses, respectively, the tuple-level attack score is defined as follows:

$$\begin{aligned} s_{\text{pair}}(z) = {} & [\bar{\ell}_\theta(y_w|x) - \bar{\ell}_{\text{ref}}(y_w|x)] \\ & -[\bar{\ell}_\theta(y_l|x) - \bar{\ell}_{\text{ref}}(y_l|x)] \end{aligned} \quad (3)$$

where $\bar{\ell}$ denotes mean response-token log-probability, $\theta$ denotes the DPO-aligned model, and ref denotes the frozen pre-DPO model. Higher scores indicate stronger evidence that the complete preference tuple was used during alignment.

A unique tuple is labeled as a member if it is sampled at least once during the 100 DPO optimization steps of a training run. Tuples available in the corresponding training pool but not sampled during these steps serve as non-member candidates. Each attack set contains equal numbers of member and non-member tuples, with the source composition of the two groups matched where possible. Membership therefore refers to the realized update stream rather than a fixed train-test split. We report tuple-level AUROC and AUPRC. On these balanced attack sets, values closer to 0.5 indicate weaker membership distinguishability. This evaluation is an empirical privacy audit rather than a formal privacy guarantee.

## IV. Experiments and Results

We compare the Baseline with privacy ratios 0.5, 1.0, and 2.0 on Gemma 3 270M-IT across five random seeds and on 4-bit Gemma 2 2B-IT across three random seeds. Both experiments evaluate helpfulness preference accuracy, harmlessness preference accuracy, and the canary suffix log-likelihood proxy. The 2B validation additionally reports PREMIA-based membership-inference AUROC and AUPRC. The canary suffix log-likelihood proxy and PREMIA metrics are interpreted as empirical privacy-risk signals rather than formal privacy guarantees.

## A. Experimental Setup

We evaluate the four configurations on Gemma 3 270M-IT across five random seeds and 4-bit-quantized Gemma 2 2B-IT across three random seeds. Table 1 summarizes the nominal training-pool composition, and Table 2 reports the model, training, and evaluation settings. For each trained 2B adapter, we construct the balanced attack sets following Section III-F. They contain 83 to 100 member tuples and an equal number of non-member tuples per run, with source composition matched where possible.

TABLE 1. TRAINING DATASET COMPOSITION

| Run | Helpfulness examples | Harmlessness examples | Privacy-preference examples | Total examples |
|---|---|---|---|---|
| Baseline | 500 | 500 | 0 | 1000 |
| Privacy ratio 0.5 | 500 | 500 | 269 | 1269 |
| Privacy ratio 1.0 | 500 | 500 | 538 | 1538 |
| Privacy ratio 2.0 | 500 | 500 | 1076 | 2076 |

TABLE 2. HYPERPARAMETERS AND REPRODUCIBILITY SETTINGS FOR THE PRIMARY AND LARGER-MODEL EXPERIMENTS

| Settings | Experimental Configuration | |
|---|---|---|
| | *270M Primary* | *2B Validation* |
| Base model | Gemma 3 270M-IT | Gemma 2 2B-IT |
| Random seeds | Five | Three |
| Base-model weights | Frozen | Frozen, MLX 4-bit |
| LoRA configuration | Rank 8; scale 10.0; dropout 0.0; four layers | Rank 8; scale 10.0; dropout 0.0; four layers |
| Optimizer and DPO | Adam optimizer; learning rate $5 \times 10^{-6}$; DPO beta 0.1 | Adam optimizer; learning rate $5 \times 10^{-6}$; DPO beta 0.1 |
| Training schedule | 100 optimization steps; maximum length 384 | 100 optimization steps; maximum length 384 |
| Evaluation set sizes | 150 helpful; 150 harmless; 100 canaries | 150 helpful; 150 harmless; 100 canaries |

## B. Evaluation Metrics

Under the evaluation protocol in Section III-E, higher preference accuracy indicates better performance on the corresponding helpfulness or harmlessness set, whereas lower canary suffix log-likelihood proxy values indicate a weaker measured memorization-related signal. For the 2B validation, tuple-level PREMIA attack performance is reported using AUROC and AUPRC, as described in Section III-F. On the balanced attack sets, values closer to 0.5 indicate weaker membership distinguishability. Results are reported as the mean ± sample standard deviation across five seeds for the 270M experiment and three seeds for the 2B validation, including the PREMIA-based metrics.

## C. Primary Gemma 3 270M-IT Results

Table 3 reports the primary results for Gemma 3 270M-IT across the non-privacy Baseline and the three privacy-aware configurations.

TABLE 3. FIVE-SEED GEMMA 3 270M-IT RESULTS (MEAN ± STANDARD DEVIATION)

| Run | Helpfulness Pref. Acc. | Harmlessness Pref. Acc. | Canary Suffix Log-likelihood Proxy |
|---|---|---|---|
| Baseline | 0.579 ± 0.009 | 0.461 ± 0.020 | -7.952 ± 1.164 |
| Privacy ratio 0.5 | 0.580 ± 0.005 | 0.441 ± 0.017 | -9.487 ± 1.164 |
| Privacy ratio 1.0 | 0.580 ± 0.012 | 0.436 ± 0.008 | -9.706 ± 1.255 |
| Privacy ratio 2.0 | 0.588 ± 0.006 | 0.427 ± 0.007 | -10.644 ± 1.162 |

All privacy-aware configurations have lower mean canary suffix log-likelihood proxy values than the Baseline (-7.952), with the lowest value observed at ratio 2.0 (-10.644). Mean helpfulness preference accuracy remains stable across configurations (0.579 to 0.588), whereas mean harmlessness preference accuracy decreases monotonically from 0.461 to 0.427. Because all harmlessness accuracies are below 0.5, these results represent relative changes under matched settings rather than evidence of strong absolute harmlessness performance. Thus, in the primary experiment, higher privacy ratios are associated with lower canary proxy values and lower harmlessness accuracy, while helpfulness accuracy remains broadly stable.

## D. Gemma 2 2B-IT Validation and Membership-Inference Results

Table 4 reports the 2B validation results across three seeds. Because the canary suffix log-likelihood proxy is model-dependent, its absolute magnitude is interpreted within this model setting rather than compared directly with the 270M results.

TABLE 4. GEMMA 2 2B-IT VALIDATION RESULTS (THREE RANDOM SEEDS)

(A) PREFERENCE ACCURACY AND CANARY SUFFIX LOG-LIKELIHOOD PROXY

| Run | Helpfulness Pref. Acc. | Harmlessness Pref. Acc. | Canary Suffix Log-likelihood Proxy |
|---|---|---|---|
| Baseline | 0.611 ± 0.021 | 0.444 ± 0.008 | -7.348 ± 1.289 |
| Privacy ratio 0.5 | 0.616 ± 0.004 | 0.431 ± 0.017 | -11.764 ± 0.084 |
| Privacy ratio 1.0 | 0.609 ± 0.017 | 0.431 ± 0.008 | -12.616 ± 1.009 |
| Privacy ratio 2.0 | 0.598 ± 0.017 | 0.489 ± 0.073 | -31.905 ± 4.869 |

(B) PREMIA-BASED MEMBERSHIP-INFERENCE RESULTS

| Run | AUROC | AUPRC |
|---|---|---|
| Baseline | 0.804 ± 0.061 | 0.790 ± 0.055 |
| Privacy ratio 0.5 | 0.608 ± 0.017 | 0.564 ± 0.014 |
| Privacy ratio 1.0 | 0.629 ± 0.025 | 0.575 ± 0.021 |
| Privacy ratio 2.0 | 0.596 ± 0.059 | 0.541 ± 0.045 |

Table 4 shows that all privacy-aware configurations yield lower mean canary suffix log-likelihood proxy values than the Baseline (-7.348). The privacy ratio 2.0 configuration yields the lowest mean value (-31.905), although it also exhibits the greatest variability across seeds. Helpfulness preference accuracy remains broadly stable across configurations, ranging from 0.598 to 0.616. Mean harmlessness preference accuracy is modestly lower at privacy ratios 0.5 and 1.0 than in the Baseline but increases to 0.489 ± 0.073 at ratio 2.0. Thus, unlike the 270M experiment, the 2B validation does not exhibit a monotonic decline in harmlessness preference accuracy as the privacy ratio increases.

On the mixed-source attack sets, the privacy-aware configurations yield lower aggregate AUROC (0.596 to 0.629) and AUPRC (0.541 to 0.575) than the Baseline (0.804 and 0.790, respectively), indicating weaker aggregate membership distinguishability. However, neither metric varies monotonically with the privacy ratio, all privacy-aware values remain above 0.5, and the aggregate pattern does not imply uniform reductions across data sources. Table 5 therefore reports the source-stratified results.

TABLE 5. SOURCE-STRATIFIED PREMIA-BASED MEMBERSHIP-INFERENCE RESULTS FOR GEMMA 2 2B-IT. VALUES ARE REPORTED AS THE MEAN ± STANDARD DEVIATION ACROSS THREE RANDOM SEEDS

(A) AUROC

| Configuration | Tuple source | | |
|---|---|---|---|
| | Helpfulness | Harmlessness | Privacy preference |
| Baseline | 0.815 ± 0.120 | 0.795 ± 0.032 | — |
| Privacy ratio 0.5 | 0.659 ± 0.063 | 0.681 ± 0.028 | 0.469 ± 0.014 |
| Privacy ratio 1.0 | 0.767 ± 0.024 | 0.737 ± 0.045 | 0.539 ± 0.106 |
| Privacy ratio 2.0 | 0.806 ± 0.081 | 0.813 ± 0.036 | 0.493 ± 0.139 |

(B) AUPRC

| Configuration | Tuple source | | |
|---|---|---|---|
| | Helpfulness | Harmlessness | Privacy preference |
| Baseline | 0.820 ± 0.103 | 0.776 ± 0.053 | — |
| Privacy ratio 0.5 | 0.627 ± 0.056 | 0.707 ± 0.032 | 0.508 ± 0.016 |
| Privacy ratio 1.0 | 0.769 ± 0.043 | 0.714 ± 0.067 | 0.560 ± 0.077 |
| Privacy ratio 2.0 | 0.817 ± 0.076 | 0.830 ± 0.024 | 0.504 ± 0.070 |

Each source-specific subset contains equal numbers of members and non-members, ranging from 17 to 55 per class across configurations and seeds. Because some subsets are small and the validation uses only three seeds, these results should be treated as descriptive source-level diagnostics. The Baseline contains no privacy-preference tuples and therefore has no corresponding source-specific result.

Ratios 0.5 and 1.0 generally have lower helpfulness and harmlessness AUROC/AUPRC values than the Baseline, whereas ratio 2.0 is comparable to or higher than the Baseline. Privacy-preference results remain near chance across the three privacy-aware configurations. Thus, the lower aggregate attack performance does not hold uniformly across sources and may partly reflect mixed-source attack-set composition. The aggregate results should therefore be interpreted as mixture-level membership distinguishability rather than a uniform source-level reduction.

Overall, the 2B validation likewise shows lower mean canary proxy values than the Baseline for all privacy-aware configurations and broadly stable helpfulness accuracy. However, it does not reproduce the monotonic decline in harmlessness accuracy observed in the primary experiment, and the source-stratified results do not show a uniform reduction in membership distinguishability.

## V. DISCUSSION

Across both model settings, privacy-preference mixing is associated with lower mean canary suffix log-likelihood proxy values. Helpfulness remains broadly stable, whereas harmlessness trends vary by model setting and the 2B membership-inference results vary by data source. These findings position P3M as a data-composition protocol for examining privacy-utility-safety trade-offs rather than as a formal privacy defense or a uniformly beneficial intervention across metrics and data sources.

### A. Interpreting the Privacy Mixing Ratio

No privacy ratio is uniformly optimal. Higher ratios correspond to lower mean canary proxy values in both models, but PREMIA attack performance in the 2B validation is non-monotonic across ratios and does not decrease uniformly across sources, while the monotonic decline in harmlessness preference accuracy observed in the 270M experiment does not recur in the 2B validation. Configurations should therefore be evaluated jointly across privacy, helpfulness, and harmlessness outcomes rather than by privacy-preference data volume alone.

With 100 optimization steps per adapter, higher ratios shift the expected update composition toward privacy-preference examples. Differences in Gemma generation, model capacity, quantization, and optimization dynamics may also contribute. Because the present experiments do not isolate these factors, this explanation remains a hypothesis.

### B. Scope and Limitations

This resource-bounded study evaluates the 270M and 2B models across five and three seeds, respectively, using LoRA-DPO with four adapted transformer layers. Because the settings differ in model generation, parameter scale, and quantization, the 2B experiment should be interpreted as a larger-model validation rather than a controlled scaling study. Future work should assess generalization to other model families and scales as computational resources permit.

The privacy ratio changes both data composition and nominal pool size because privacy-preference rows are added or repeated rather than substituted for non-privacy rows. Although all configurations use 100 optimization steps, the inclusion of an example in the nominal pool does not guarantee that it is sampled during training. Thus, the observed metric differences may reflect not only privacy-preference data composition but also pool size, duplicated privacy rows, and variation in the examples sampled across runs. A fixed-total-size control that replaces some non-privacy examples with privacy-preference examples would help isolate composition effects from changes in pool size.

The canary proxy and PREMIA-based evaluation provide complementary empirical diagnostics of memorization-related signals and membership distinguishability. They do not provide formal privacy guarantees or establish resistance to extraction attacks. In our PREMIA-based evaluation, membership is defined with respect to the realized update stream rather than a fixed train-test partition. Because the balanced attack sets vary across runs in size and source composition, aggregate comparisons may be sensitive to attack-set composition. Future work should evaluate fixed member and non-member sets across runs, additional membership-inference protocols, and generation-based extraction attacks.

Synthetic canaries avoid the need for real personal data but do not capture leakage involving realistic identifiers, contextual clues, or combinations of partial information. More representative privacy-sensitive formats and extraction prompts should therefore be evaluated under appropriate safeguards.

## VI. Conclusion

This paper introduces P3M, a DPO data-composition protocol that varies the amount of synthetic privacy-preference data while keeping the helpfulness and harmlessness pools fixed within each model experiment. Under the tested conditions, all privacy-aware configurations show lower mean canary suffix log-likelihood proxy values than the corresponding Baseline within each model setting. In the PREMIA-based evaluation, the privacy-aware 2B configurations also yield lower aggregate AUROC and AUPRC values than the Baseline on the constructed mixed-source attack sets, although the source-stratified results do not show a uniform reduction in membership distinguishability across data sources. Helpfulness preference accuracy remains broadly stable, whereas harmlessness trends differ between the two model settings. In similar resource-bounded settings, privacy-mixing configurations should therefore be evaluated jointly across privacy, utility, and safety outcomes rather than selected on the basis of any single objective. P3M should be viewed as a lightweight empirical protocol for examining privacy-utility-safety trade-offs rather than as a formal privacy guarantee or evidence of resistance to extraction attacks. Future work should evaluate fixed-total-size controls, additional model families, more realistic privacy-sensitive formats, and broader membership-inference and extraction evaluations.